# Title page

Names of the authors: F. Tárkányi[a], F. Ditrói[a], S. Takács[a], A. Hermanne[b], B. Király[a]

Title: Activation cross-section data for alpha-particle induced nuclear reactions on natural ytterbium for some longer lived radioisotopes

Affiliation(s) and address(es) of the author(s):

[a]Institute for Nuclear Research, Hungarian Academy of Sciences (ATOMKI), 4026 Debrecen, Bem ter 18/c, Hungary
[b] Cyclotron Laboratory, Vrije Universiteit Brussel (VUB), 1090 Brussels, Belgium

E-mail address of the corresponding author: ditroi@atomki.hu

# Activation cross-section data for alpha-particle induced nuclear reactions on natural ytterbium for some longer lived radioisotopes

F. Tárkányi[a], F. Ditrói[a1], S. Takács[a], A. Hermanne[b], B. Király[a]

[a] Institute for Nuclear Research, Hungarian Academy of Sciences (ATOMKI), 4026 Debrecen, Bem ter 18/c, Hungary

[b] Cyclotron Laboratory, Vrije Universiteit Brussel (VUB), 1090 Brussels, Belgium

## Abstract

Additional experimental cross sections were deduced for the long half-life activation products ($^{172}$Hf and $^{173}$Lu) from the alpha particle induced reactions on ytterbium up to 38 MeV from late, long measurements and for $^{175}$Yb, $^{167}$Tm from a re-evaluation of earlier measured spectra. The cross-sections are compared with the earlier experimental datasets and with the data based on the TALYS theoretical nuclear reaction model (available in the TENDL-2014 and 2015 libraries) and the ALICE-IPPE code.

[1] Corresponding author: ditroi@atomki.hu

## Introduction

In the frame of our systematic investigation of production routes of radioisotopes used in medicine for radiotherapy, we published experimental excitation functions of α–particle induced nuclear reactions on natural Yb target for the nuclear reactions $^{nat}Yb(\alpha,xn)^{170,171,173,175,177m2}Hf$, $^{nat}Yb(\alpha,x)^{171g,172g,177g,178m}Lu$ and $^{nat}Yb(\alpha,x)^{169g,177g}Yb$ [1]. As the samples were measured after up to 2 weeks cooling times, with measuring times of 1-2 h, some long half-life isotopes were not detected.

When re-measuring the gamma-spectra of these ytterbium samples for long time (20-28 h) after 1 year of cooling, still statistically significant additional activities were detected making possible to deduce cross-sections for some longer-lived radio-products. The γ-ray signals from these long-lived radioisotopes were masked in the previous, shorter, measurements by the "background" (Compton scattering, multiple low abundance γ-rays, ….) generated by the shorter half-life isotopes. During the assessment of the new spectra, also the earlier measurements were re-evaluated and presence of some not reported, shorter-lived activation products were noted ($^{175}Yb$, $^{167}Tm$). We thought, it is meaningful to report all these additional production cross-section data, for which no or only one data set, published by Romo et al. [2], exists.

## Experimental procedure and the data processing

The activation cross sections were measured by using the stacked foil irradiation technique and high resolution gamma spectrometry. Two stacks containing Ti and Yb foils were irradiated (referred as stack1 and stack2) with 40 MeV alpha particles. The two stacks were shifted in energy by using energy degrader foils. The $^{nat}Ti$ monitors [3,4] and the $^{nat}Yb$ target foils were measured directly, without chemical separation. The details of the experimental procedure and the data processing are described in the previous reports [1,5-8]. The used decay data and the Q-values of contributing processes are collected in Table 1 [9,10]. As ytterbium has seven stable isotopes so called elemental cross-sections are deduced.

Table 1 Decay data of the investigated reaction products

| Nuclide<br>Decay path | Half-life | $E_\gamma$(keV) | $I_\gamma$(%) | Contributing reactions | Q-value (keV) GS-GS |
|---|---|---|---|---|---|
| **$^{172}$Hf**<br>ε: 100 % | 1.87 a | 125.81 | 11.3 | $^{170}$Yb(α,2n)<br>$^{171}$Yb(α,3n)<br>$^{172}$Yb(α,4n)<br>$^{173}$Yb(α,5n)<br>$^{174}$Yb(α,6n) | -18080.2<br>-28123.2<br>-32714.3<br>-44246.3<br>-52752.5 |
| **$^{173}$Lu**<br>ε: 100 % | 1.37 a | 272.11 | 21.2 | $^{170}$Yb(α,p)<br>$^{171}$Yb(α,pn)<br>$^{172}$Yb(α,p2n)<br>$^{173}$Yb(α,p3n)<br>$^{174}$Yb(α,p4n)<br>$^{176}$Yb(α,p6n)<br>$^{173}$Hf decay | -8746.2<br>-15360.8<br>-23380.3<br>-29747.7<br>-37212.3<br>-49899.1 |
| **$^{175}$Yb**<br>$\beta^-$: 100 % | 4.185 d | 282.52<br>396.33 | 6.13<br>13.2 | $^{173}$Yb(α,2p)<br>$^{174}$Yb(α,2pn)<br>$^{176}$Yb(α,2p3n)<br>$^{175}$Tm decay | -15008.7<br>-22473.3<br>-35160.0 |
| **$^{168}$Tm**<br>ε: 99.99%<br>$\beta^-$: 0.010 % | 93.1 d | 198.25<br>815.99 | 54.49<br>50.95 | $^{168}$Yb(α,3pn)<br>$^{170}$Yb(α,3p3n)<br>$^{171}$Yb(α,3p4n)<br>$^{172}$Yb(α,3p5n)<br>$^{173}$Yb(α,3p6n)<br>$^{174}$Yb(α,3p7n) | -27781.4<br>-43107.3<br>-49722.0<br>-57741.4<br>-64108.8<br>-71573.4 |
| **$^{167}$Tm**<br>ε: 100 % | 9.25 d | 207.80 | 42 | $^{168}$Yb(α,3p2n)<br>$^{170}$Yb(α,3p4n)<br>$^{171}$Yb(α,3p5n)<br>$^{172}$Yb(α,3p6n)<br>$^{173}$Yb(α,3p7n)<br>$^{167}$Yb decay | -34622.0<br>-49948.0<br>-56562.6<br>-64582.0<br>-70949.4 |

When complex particles are emitted instead of individual protons and neutrons the Q-values have to be decreased by the respective binding energies: pn→d +2.2 MeV, p2n→t +8.5 MeV, 2pn→$^{3}$He +7.7 MeV, 2p2n→α +28.3 MeV
Abundances in $^{nat}$Yb: $^{1168}$Yb 0.13%, $^{170}$Yb 3.45%,$^{171}$Yb 14.3%, $^{172}$Yb 21.9%, $^{173}$Yb 16.12%, $^{174}$Yb 31.8%, $^{176}$Yb 12.7%

## Results

Activation cross-section data were determined from the new measurements for production of long-lived $^{172}$Hf and $^{173}$Lu and from the re-evaluation of the previous spectra for $^{175}$Yb and $^{167}$Tm. The cross sections are shown in Figures 1-4 in comparison with the only earlier experimental data published by Romo et al. [2] and with the theoretical results in the TENDL-2014, 2015 libraries [11], based on TALYS code [12] as well as with the results of the ALICE-IPPE code [13,14]. The numerical cross-section values are collected in Table 2.

*$^{172}$Hf*

The measured activation cross sections of $^{172}$Hf ($T_{1/2}$ = 1.87 a, ε: 100 %, $J^{\pi} = 0^{+}$) are shown in Fig. 1. The $^{172}$Hf is produced directly via (α,xn) reactions. The agreement with the experimental data of Romo et al [2] and with the theoretical data is acceptable, both as the trend and as the values regarded. There is a difference by some points between our two experiments (stacks), which can be explained by the fact that in spite of the long measuring times the statistics of the peak areas is still relatively poor.

*$^{173}$Lu*

The $^{173}$Lu ***(***1.37 a) cross-sections are cumulative and include direct production by (α,pxn) reactions and decay of the shorter-lived $^{173}$Hf (23.6 h, ε: 100 %) parent. Our results are in good agreement with the earlier data of Romo et al. [1], but a better agreement is seen with our stack 1 data above 25 MeV and with our stack 2 data under this energy. The theoretical data are systematically lower, probably because of underestimation of the direct production where a proton is emitted (Fig. 2).

*$^{175}$Yb*

The cross-sections of the $^{175}$Yb (4.185 d) are cumulative, including direct production by $^{nat}$Yb(α,2pxn) reactions and the decay through short-lived $^{175}$Tm parent (15.2 min, $\beta^{-}$: 100 %). Our data, obtained from the re-evaluation of earlier spectra, are in acceptable agreement with earlier results of Romo et al. [1]. The theory significantly underestimates the experimental values and gives a curve with different shape (Fig. 3).

### *$^{168}$Tm*

The $^{168}$Tm (93.1 d, $3^+$, ε: 99.99 %, $\beta^-$: 0.010 %) is produced directly via (α,3pxn) reactions, with a predominant contribution of clustered emissions in the energy domain studied. We could measure some gamma-lines of $^{168}$Tm with poor statistics and contradicting values, the experimental data we could deduce were very scattered and far from the expected range, that's why it is not presented in this work.

### *$^{167}$Tm*

The $^{167}$Tm (9.25 d, ε: 100 %) is produced directly and through the $^{167}$Yb parent (17.5 min, ε: 100 %) decay. The experimental data, derived from the re-evaluation of the earlier spectra, are higher compared to the theoretical data in the TENDL libraries. (Fig. 4). ALCE-IPPE gives an even lower estimation. Both theoretical nuclear model code calculation results follow the trend of the experimental data.

Table 2 Cross sections of the measured isotopes

| Alpha energy E ± ΔE (MeV) | | $^{172}$Hf | | $^{173}$Lu | | $^{175}$Yb | | $^{167}$Tm | |
|---|---|---|---|---|---|---|---|---|---|
| | | Cross section σ ± Δσ | | | | | | | |
| 37.7 | 0.3 | 190.7 | 21.6 | 573.1 | 64.5 | 10.9 | 3.5 | | |
| 35.3 | 0.4 | 195.8 | 22.1 | 517.0 | 58.2 | 6.72 | 1.84 | | |
| 32.8 | 0.4 | 172.8 | 20.3 | 461.7 | 52.0 | 3.48 | 2.47 | | |
| 30.1 | 0.5 | 88.9 | 10.1 | 369.7 | 41.7 | | | | |
| 27.3 | 0.6 | 29.9 | 3.6 | 296.9 | 33.4 | | | | |
| 24.3 | 0.7 | 17.9 | 2.4 | 229.4 | 25.9 | | | | |
| 21.0 | 0.8 | | | 122.2 | 13.9 | | | | |
| 17.4 | 1.0 | | | 52.5 | 5.9 | | | | |
| 13.1 | 1.2 | | | 27.6 | 3.2 | | | | |
| 7.8 | 1.4 | | | 8.3 | 1.2 | | | | |
| | | | | | | | | | |
| 37.2 | 0.3 | 198.7 | 22.5 | 576.0 | 64.7 | 13.6 | 2.2 | 0.38 | 0.16 |
| 34.8 | 0.4 | 162.0 | 18.4 | 466.4 | 52.5 | 8.96 | 1.93 | 0.36 | 0.18 |
| 32.3 | 0.4 | 119.2 | 13.5 | 368.2 | 41.4 | 4.17 | 0.92 | 0.37 | 0.09 |
| 29.6 | 0.5 | 59.8 | 6.8 | 243.4 | 27.3 | 1.39 | 0.93 | 0.32 | 0.06 |
| 26.7 | 0.6 | 14.1 | 4.1 | 176.1 | 19.8 | | | 0.19 | 0.07 |
| 23.7 | 0.7 | 11.4 | 1.7 | 108.5 | 12.2 | | | | |
| 20.3 | 0.8 | 1.53 | 0.81 | 21.7 | 2.5 | 0.62 | 0.23 | | |

## Summary

Experimental cross sections were deduced for alpha particle induced reactions on ytterbium for $^{172}$Hf and $^{173}$Lu and up to 38 MeV from late, long spectra measurements and for $^{175}$Yb and $^{167}$Tm from a re-evaluation of earlier measured spectra. The new experimental data support the earlier data of Romo et al. [1] for $^{172}$Lu, $^{173}$Lu and $^{175}$Yb. No earlier data were found in the literature for $^{168}$Tm and $^{167}$Tm. The theoretical results based on the TALYS nuclear reaction model code (in TENDL-2014 and 2015 libraries) are significantly lower compared to the experiment (except for $^{172}$Hf and $^{173}$Lu). The ALICE-IPPE gives different results in the case of Tm radioisotopes but very similar to the TENDL calculations in the case of others. TENDL-2015 does not provide remarkable improvements compared to TENDL-2014.

**Figures**

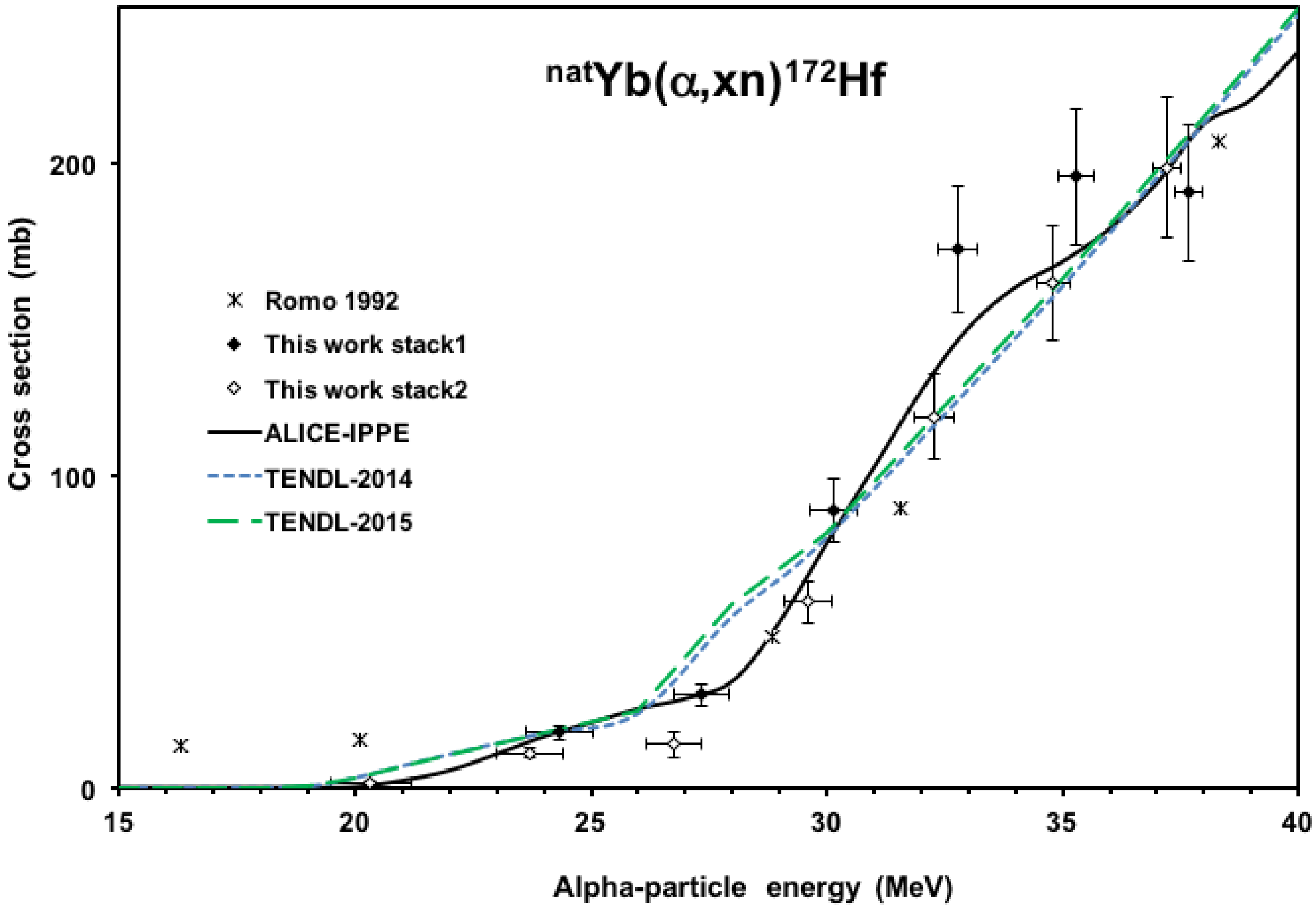

Fig. 1. Experimental and theoretical excitation functions for $^{nat}Yb(\alpha,xn)^{172}Hf$

reaction

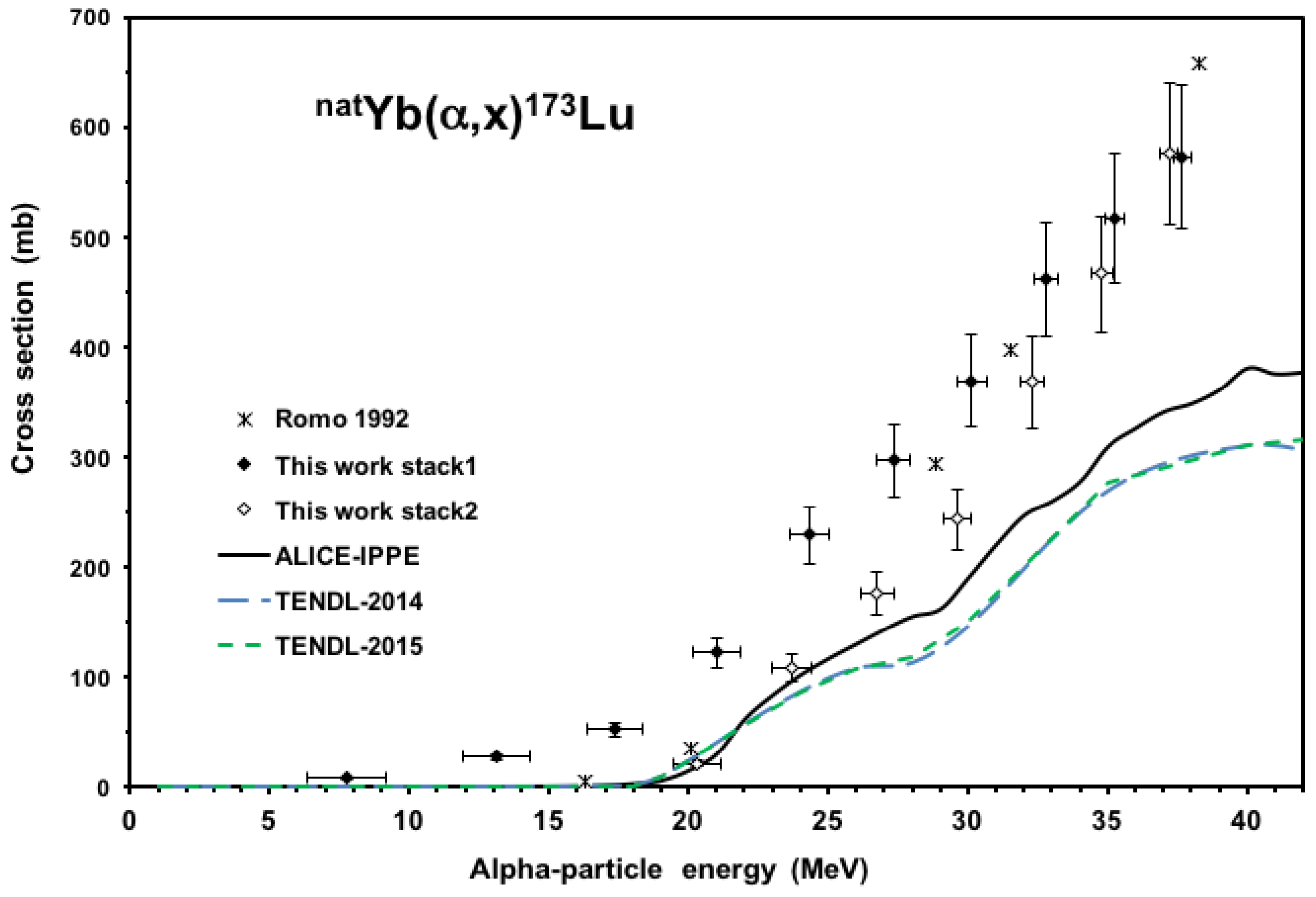

Fig. 2. Experimental and theoretical excitation functions for $^{nat}Yb(\alpha,x)^{173}Lu$ reaction

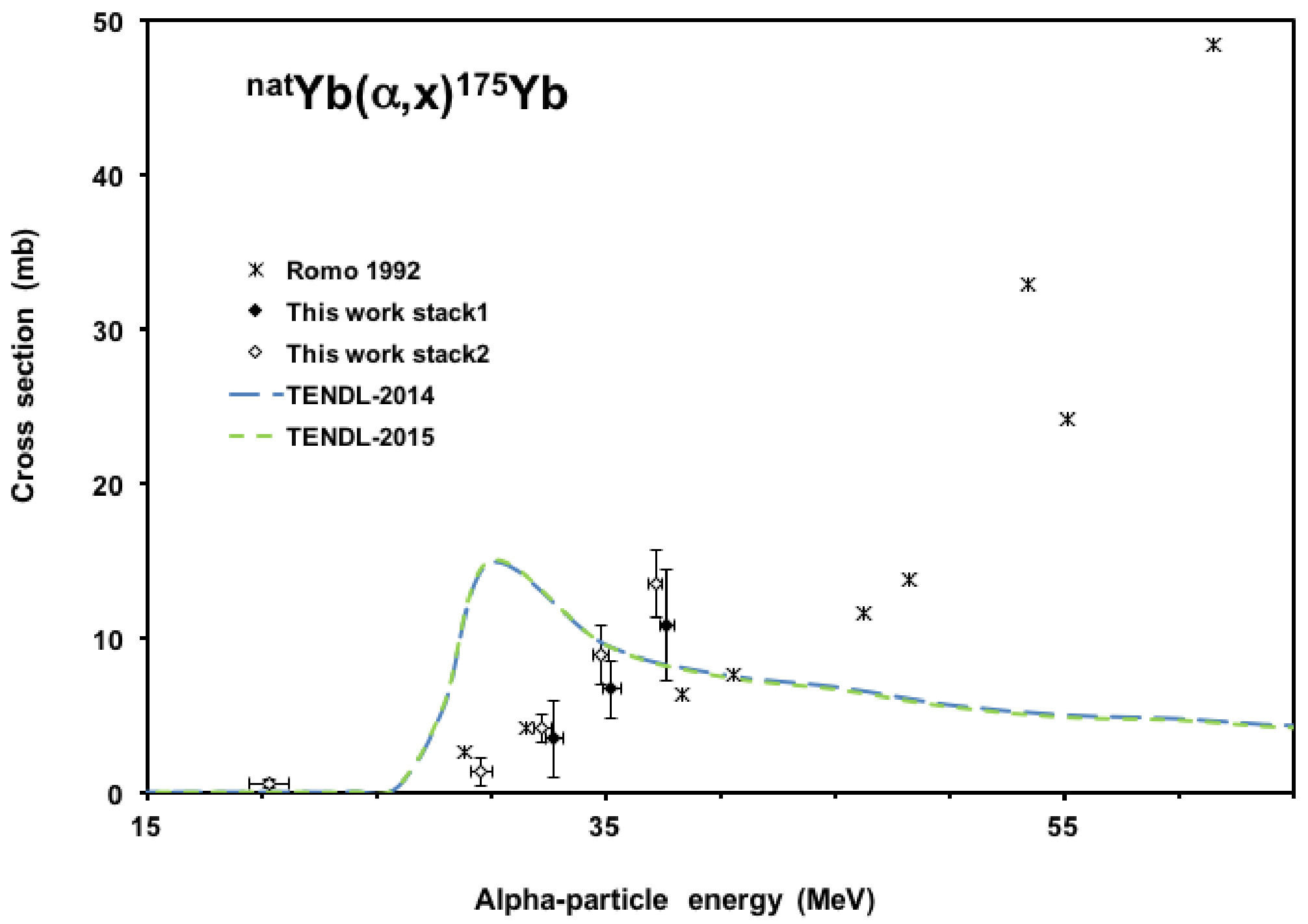

Fig. 3. Experimental and theoretical excitation functions for $^{nat}Yb(\alpha,x)^{175}Yb$ reaction

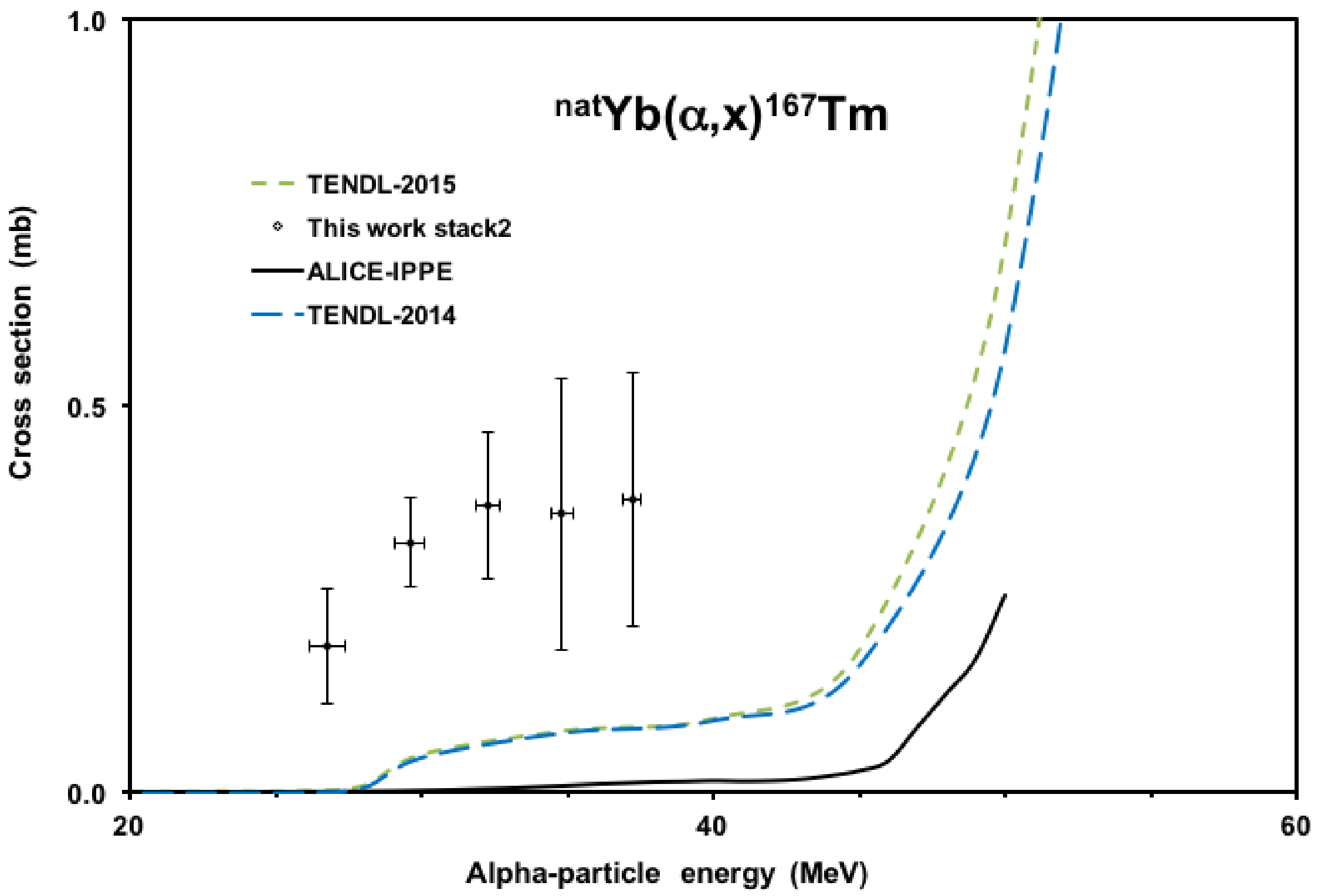

Fig. 4. Experimental and theoretical excitation functions for $^{nat}Yb(\alpha,x)^{167}Tm$ reactio*n*